\documentclass[doublecol]{epl2}
\usepackage{graphicx}
\usepackage{amssymb}
\usepackage{amsmath}
\usepackage{color}

\providecommand{\st}[1]{_{\text{#1}}}

\def\const{\mathrm{const}}
\def\deg{^\circ}

\def\sLV{\sigma\st{LV}}

\def\cap{\st{cap}}

\def\sph{\st{sph}}

\def\eff{\st{eff}}
\def\tot{\st{tot}}
\def\colspc{\hspace*{0.7cm}}

\newcommand{\bitem}{\begin{itemize}}
\newcommand{\eitem}{\end{itemize}}
\newcommand{\benum}{\begin{enumerate}}
\newcommand{\eenum}{\end{enumerate}}
\newcommand{\btab}[1]{\begin{tabular}{#1}}
\newcommand{\etab}{\end{tabular}}
\newcommand{\btabn}[1]{\begin{tabular}{#1}}
\newcommand{\etabn}{\end{tabular}}
\newcommand{\beq}{\begin{equation}}
\newcommand{\eeq}{\end{equation}}
\newcommand{\beqn}{\begin{equation*}}
\newcommand{\eeqn}{\end{equation*}}
\newcommand{\bsplit}{\begin{split}}
\newcommand{\esplit}{\end{split}}

\title{Fall and rise of small droplets on rough hydrophobic substrates}
\author{M. Gross\inst{1,2} \and F. Varnik\inst{1,2} \and D. Raabe\inst{2}}
\institute{
  \inst{1} Interdisciplinary Centre for Advanced Materials Simulation (ICAMS), Ruhr-Universit\"at Bochum, Stiepeler Strasse 129, 44801 Bochum, Germany\\
  \inst{2} Max-Planck Institut f\"ur Eisenforschung, Max-Planck Str.~1, 40237 D\"usseldorf, Germany
}
\shortauthor{M. Gross, F. Varnik, D. Raabe}


\abstract{ Liquid droplets on patterned hydrophobic substrates are typically observed either in the Wenzel or the Cassie state. Here we show that for droplets of comparable size to the roughness scale an additional local equilibrium state exists, where the droplet is immersed in the texture, but not yet contacts the bottom grooves. Upon evaporation, a droplet in this state enters the Cassie state, opening the possibility of a qualitatively new self-cleaning mechanism. The effect is of generic character and is expected to occur in any hydrophobic capillary wetting situation where a spherical liquid reservoir is involved. }

\begin{document}

\maketitle

\section{Introduction}

The fact that roughness at the micrometer level can drastically increase the water-repellant properties of a hydrophobic substance has been known for a long time \cite{wenzel_resistance_1936, cassie_wettability_1944} and is a frequent phenomenon in nature -- the most prominent example being the Lotus leaf. Technical advancement in surface fabrication and novel industrial applications, such as self-cleaning materials, has brought the phenomenon of superhydrophobicity again in the focus of research during the last years (see \cite{qur_wetting_2008, dorrer_thoughts_2009} for recent reviews).

Until now, mostly droplets that are much larger than the typical roughness scale of the surface have been investigated. In that case, the droplet either appears in the Wenzel state, where it completely wets the substrate~\cite{wenzel_resistance_1936}, or in the Cassie state, i.e.\ on top of the roughness structures~\cite{cassie_wettability_1944}.
The characteristic low-adhesion, water-repellant properties of superhydrophobic surfaces are associated with droplets in the Cassie state. Contrarily, the Wenzel state leads to sticky, highly pinned droplets \cite{lafuma_superhydrophobic_2003}. 

The fact that usually both the Cassie and the Wenzel state can be observed on the same substrate implies that both are separated by a free energy barrier, which can be overcome by external forces or kinetic energy \cite{bico_pearl_1999, lafuma_superhydrophobic_2003, he_multiple_2003}. It is known that the transition from the Cassie to the Wenzel state proceeds through the nucleation of contact between the liquid and the substrate at the grooves~\cite{sbragaglia_spontaneous_2007}, initiated e.g.\ via the increase of internal droplet pressure during evaporation \cite{jung_wetting_2007, moulinet_life_2007, reyssat_impalement_2008}. It is unclear, however, how deep a droplet can really sink into the texture and how this process is modified for droplets of similar size as the roughness scale.

The validity and applicability of the classical two-state picture of Wenzel and Cassie has been questioned several times \cite{swain_contact_1998, wolansky_apparent_1999, gao_wrong_2007}. In particular, for the case of droplets that are of comparable size to the surface roughness -- and which therefore directly feel the influence of the surface geometry -- it can not be expected to hold a priori. Due to the fact that such sub-micron-sized droplets have a very short lifetime, only few experimental hints on their behaviour have been provided so far \cite{lau_superhydrophobic_2003, jopp_wetting_2004, reyssat_impalement_2008}. Recent studies, focusing explicitly on this situation, confirmed at least the existence of the two equilibrium states \cite{jopp_wetting_2004, koishi_coexistence_2009}.

Nevertheless, droplets of this length scale are not only important for a better understanding of the wetting properties of microscale systems~\cite{seemann_wetting_2005}, but also in many industrial applications, as, for example, in the production of efficient self-cleaning surfaces \cite{blossey_selfcleaning_2003}, robust metal coatings, or in plasma spraying techniques. Moreover, since these droplets naturally occur in any condensation or evaporation process \cite{narhe_nucleation_2004, dorrer_condensation_2007}, their phenomenology is also fundamental for a better understanding of the water-repellent properties of many plant-leaves or insect eyes and legs.

In this letter, we investigate the behaviour of droplets in three dimensions both analytically and via numerical lattice Boltzmann (LB) computer simulations. We show that, for a droplet of comparable size to the surface roughness, besides the possible Wenzel and Cassie states, there exists a further generic state (hereafter called \emph{impaled state}), characterized by almost complete immersion of the droplet into the texture, yet not touching the bottom of the grooves. To the best of our knowledge, such a state has not been reported before. It is important to realize that this state is different from the ``partially impaled'' conformations of the Cassie state, where the liquid-vapor interface below the macroscopic contact line is curved, but the droplet essentially remains on top of the asperities \cite{moulinet_life_2007, kusumaatmaja_collapse_2008}. This statement is supported by the fact that we indeed find a coexistence regime for the new impaled state and the (partially impaled) Cassie state. By virtue of this new state, a droplet can in fact be saved from penetrating into the texture completely and going over to the Wenzel state. Instead, it can reenter a Cassie state upon, e.g., evaporation.
We demonstrate the generic character of this finding, which has important implications for the wetting behaviour of droplets and for the effectiveness of self-cleaning surfaces.

\section{Model}

\begin{figure}[t]
    (a)\includegraphics[width=0.35\linewidth]{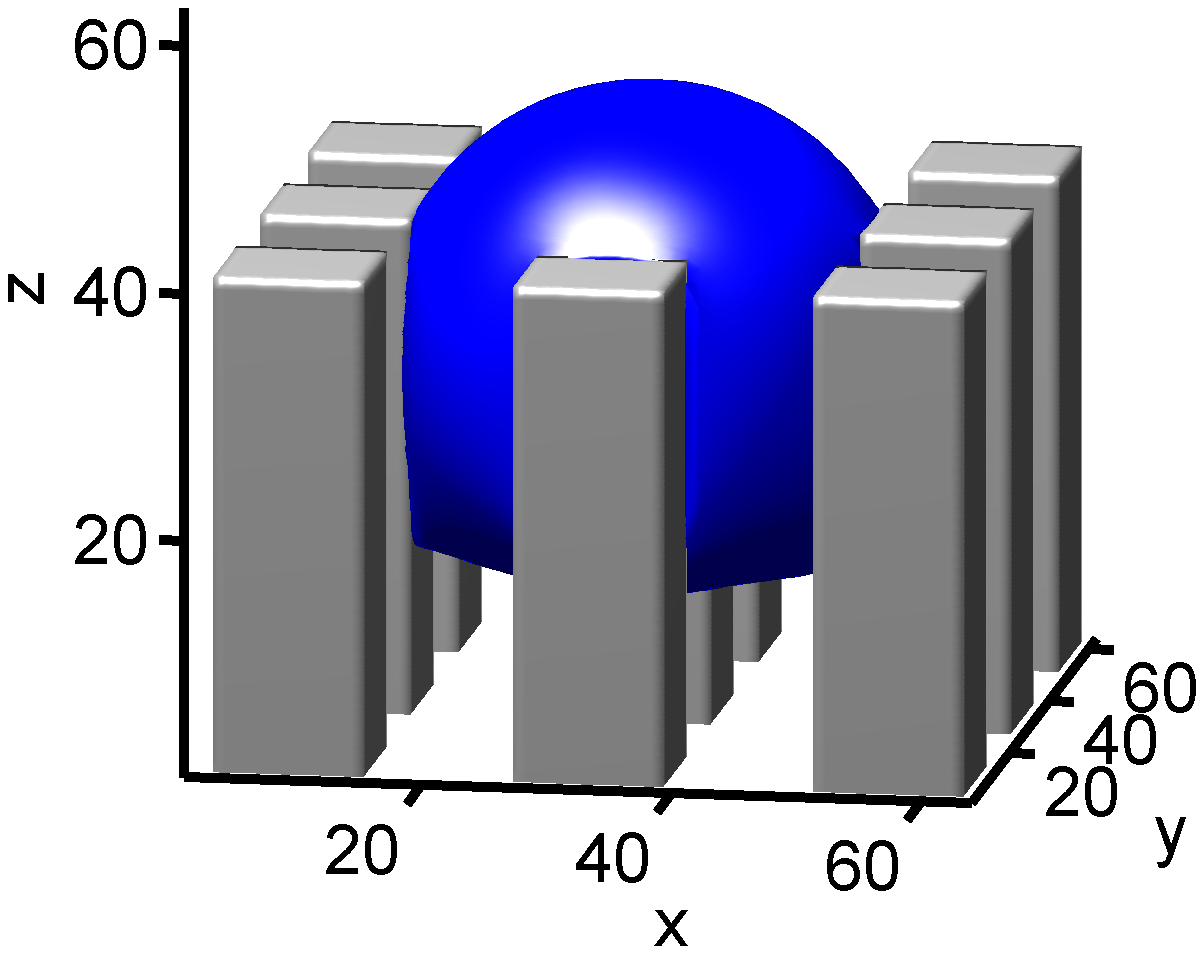} \colspc
    (b) \includegraphics[width=0.37\linewidth]{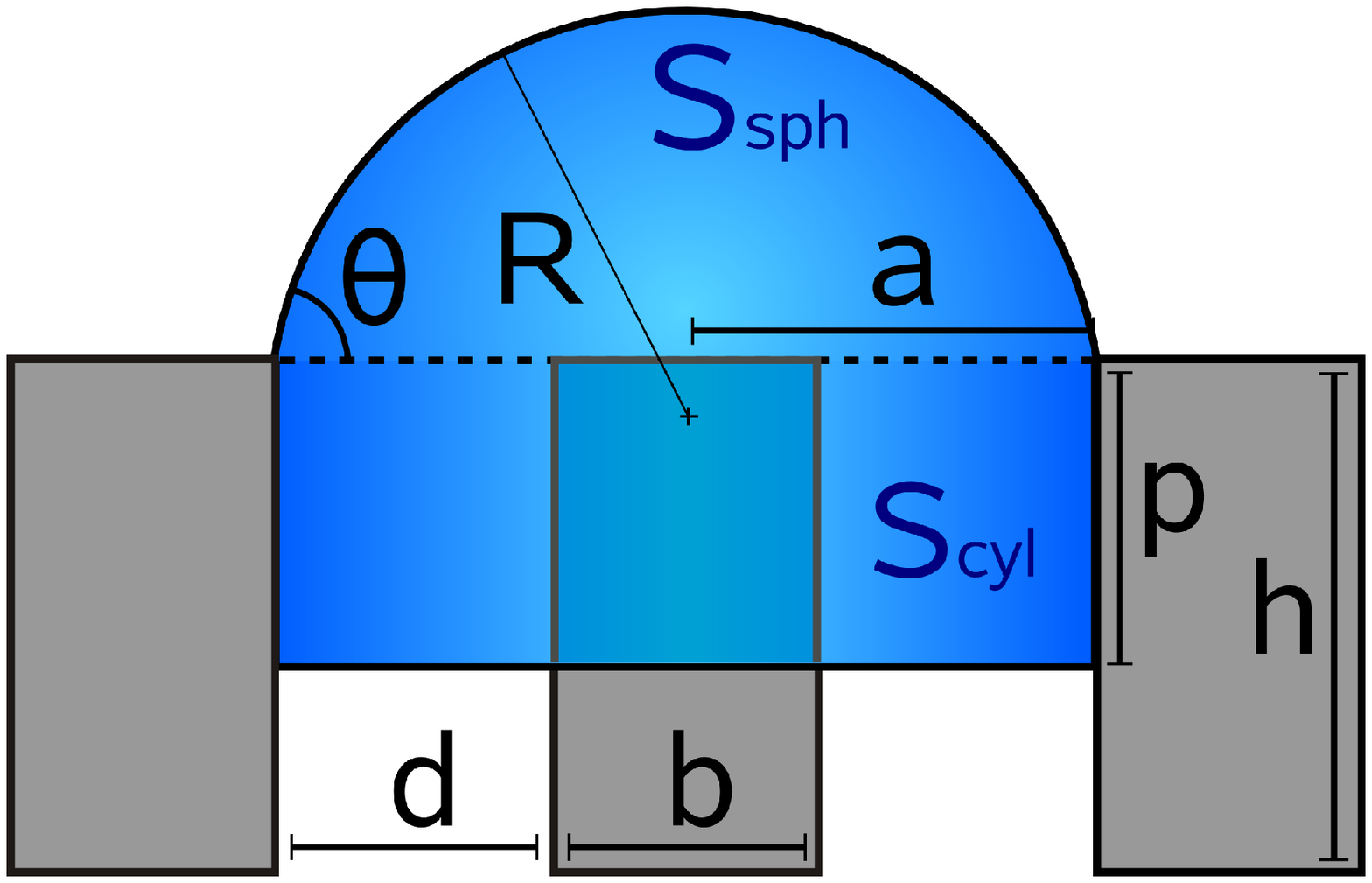}
   \caption{A small droplet on a hydrophobic pillar array. (a) shows the impaled state as a typical situation in the present simulations. In the analytical model (b), the droplet is assumed to be pinned at the edges of the pillars, i.e.\ it has a fixed base radius $a$. For a given droplet size, the only way to minimize the overall free energy is thus through a change in the penetration depth $p$. This also fixes the apparent contact angle $\theta$.}
    \label{fig:droplet-mod}
\end{figure}

The roughness of a surface is modeled by a regular array of cuboidal pillars with width $b$, height $h$ and spacing $d$ (Fig.~\ref{fig:droplet-mod}). The intrinsic hydrophobicity of the flat parts of the surface is described by the Young contact angle $\theta_Y$. In this work we only focus on the case $\theta_Y>90\deg$, which also is a necessary condition for the existence of a Cassie state. Gravity will be neglected throughout, as we only consider droplets that are smaller than the capillary length ($\sim$2.7\;mm for water).

In the analytical model, we assume the part of the droplet located above the pillars to be a spherical cap with base $a=b/2+d=R \sin\theta$ (with $R$ being the radius of the cap and $\theta$ the apparent contact angle).
The impaled part is approximated as a cylindrical liquid column with radius $a$ and height $p$ (penetration depth), surrounding the central pillar. 
The macroscopic contact line of the droplet is assumed to remain pinned at the edges of the outer pillars.
Note that in this model the \textit{Wenzel} state would correspond to $p=h$ and the \textit{Cassie} state to $p=0$.
Therefore, the model also neglects a possible finite penetration depth of a droplet in a ``partially impaled'' Cassie state.
As further simulations~\cite{moradi_morphological_2009} have shown, the symmetric droplet configuration considered here is stable against moderate perturbations.

Since the mechanism of the Wenzel transition has been discussed in detail in previous publications, we will ingore the Wenzel state completely, and, for the rest of this work, assume the pillars to be so tall that no contact between the liquid and the bottom of the grooves is possible.

The total volume of the droplet shall be fixed, $V\tot=\const=V\st{sph}(\theta)+V\st{cyl}(p)$, with 
$V\st{sph}=\frac{1}{3}\pi a^3 (2-3\cos\theta+\cos^3\theta)/\sin^3\theta$ 
the volume of the cap and \mbox{$V\st{cyl}=(\pi a^2-b^2)p$} the volume of the penetrating cylinder. In the following, instead of the droplet volume $V\tot$, we will usually refer to the effective droplet radius $R\eff$ that corresponds to a spherical droplet of the same volume ($4\pi/3R^3_{\text{eff}}=V_{\text{tot}}$). We consider $p$ as the free variable and determine the dependence of $\theta$ on $p$ via the fixed volume condition.

Since the volume of the drop (and the temperature) is constant, only surface energy contributions play a role for a change in the total free energy. The free energy $f(p)$ of the model droplet, neglecting gravity and terms associated with the Wenzel transition, and normalizing such that $f(0)=0$, is then given by
\begin{equation}
f(p)=\sLV \Bigl[ S\st{sph}(p) -S\st{sph}(0)+ S\st{cyl,LV}(p) - 8bp\cos\theta_Y \Bigr]
\,.
\label{eq:freeE}
\end{equation}
Here, $S\st{sph}=2\pi a^2 (1-\cos\theta)/\sin^2\theta$ is the surface area of the spherical cap $S\st{cyl,LV}=(2\pi a-4b) p$ is the lateral liquid-vapour surface area of the cylinder and $-\sLV 8bp\cos\theta_Y$ is the (positive, since $\theta_Y>90\deg$) free energy associated with the wetting of the (eight) side walls of the pillars.

\section{Analytical results}

\begin{figure*}[t]
    (a) \includegraphics[width=0.29\linewidth]{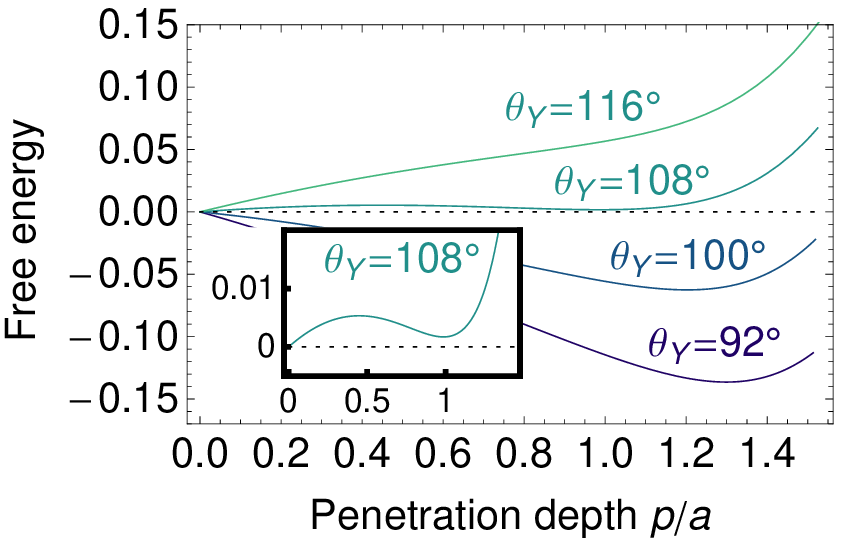} \;
    (b) \includegraphics[width=0.29\linewidth]{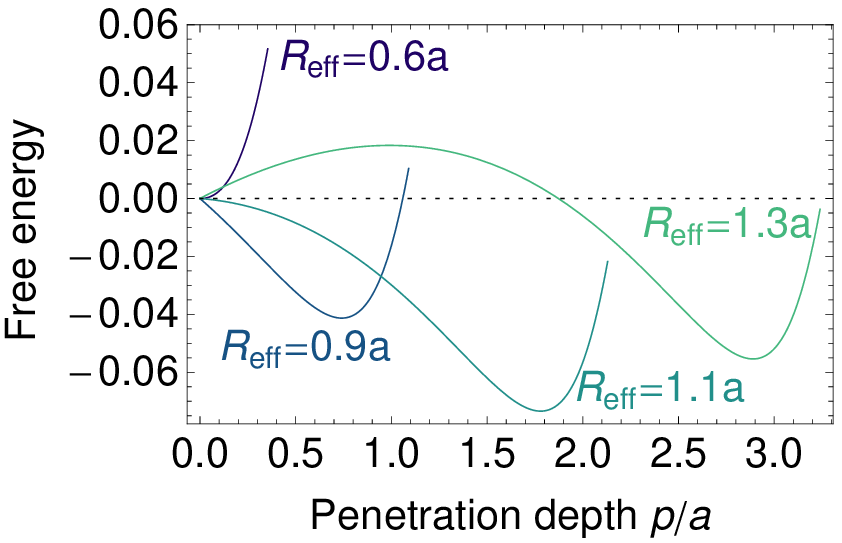} \;
    (c) \includegraphics[width=0.29\linewidth]{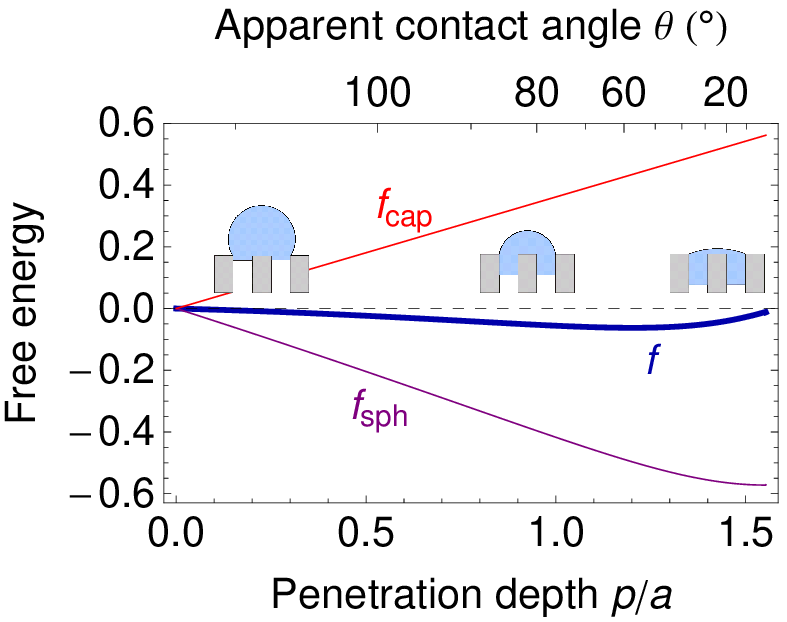} \;
    \caption{Predictions of the analytical model. (a) Dependence of the free energy $f/4\pi a^2\sLV$ on the penetration depth $p$ and the Young contact angle $\theta_Y$ for a droplet of fixed size $R\eff=a$. (The inset shows a magnification of the curve for $\theta_Y=108\deg$.)
    (b) Dependence of the free energy $f/4\pi a^2 \sLV$ on the penetration depth $p$ and the droplet size $R\eff$ for a fixed contact angle of $\theta_Y=100\deg$.
    (c) Contributions to the free energy $f/4\pi a^2 \sLV$ for $R\eff=a$ and $\theta_Y=100\deg$. $f\st{sph}$ is the surface free energy of the liquid-vapour interface of the spherical cap, $f\st{cap}$ is the free energy due to the wetting of the ``capillary'' constituted by the pillars and $f=f\sph+f\cap$ is the total free energy. The insets sketch the droplet configuration for different $p$ according to the analytical model.
All curves in (a-c) are given for $b/d=1$ and plotted up to a value of $p$ where no further volume is left in the spherical cap.
}
    \label{fig:model-predict}
\end{figure*}

Figures~\ref{fig:model-predict}a,b show the dependence of the free-energy on the penetration depth $p$ for varying Young contact angles and droplet sizes. Several interesting observations can be made: Firstly, as also found in the case of droplets large compared to the roughness scale~\cite{lafuma_superhydrophobic_2003, mchale_analysis_2005, moulinet_life_2007, reyssat_impalement_2008, kusumaatmaja_collapse_2008}, the stability of the Cassie state, determined by the slope of $f$ at $p=0$, depends not only on the contact angle but also on the size of the droplet.

The novel feature is the appearance of a \textit{local minimum} of the free energy at large penetration depths, existing \textit{in addition} to the possible minimum associated with the Cassie (and Wenzel) state.
From the condition \mbox{$\upd f/ \upd p=0$}, which is easily evaluated with the help of the fixed volume constraint,
a necessary condition for the existence of a minimum of the free energy arises, namely, \mbox{$\theta_Y < \arccos(-\frac{1}{2}+\frac{b}{4a})$}, with $a=b/2+d$ being the base radius of the spherical cap. Interestingly, this condition does not depend on the droplet size.

The origin of this new state can be understood by imagining the pillars to represent a (partly open) hydrophobic capillary tube, wetted by a small droplet that is placed at its entry. In this situation, the equilibrium state of the droplet is a consequence of the balance between the Laplace pressure within the spherical cap (pushing the droplet into the capillary) and an opposing capillary force due to the hydrophobicity of the substrate.

To illustrate this idea, we split the free energy [Eq.~\eqref{eq:freeE}] into the contributions of the spherical cap and the remaining ``capillary'' part. As shown in (Fig.~\ref{fig:model-predict}c), an increase of the droplet penetration $p$ leads to a  linear increase of capillary free energy, while the free energy associated with the spherical cap decreases in a non-linear fashion. As a result, the total free energy $f$ may exhibit a local minimum. This simple reasoning suggests that the intermediate minimum constitutes a generic equilibrium state of a droplet, occurring in any situation of filling hydrophobic capillaries by a spherical liquid reservoir. Indeed, further simulations using various surface geometries (e.g.\ omitting the central pillar) clearly underline this assertion \cite{moradi_morphological_2009}.

These results also show under which conditions we are allowed not to consider the Wenzel state in the first place: A transition to this state can be inhibited, if the pillar height $h$ is larger than the penetration depth $p$ of a droplet at the local minimum, plus a small correction of the order of $d^2/R\eff$ \cite{qur_wetting_2008} that accounts for the curvature of the lower droplet interface.

\begin{figure}[t]
    \centering
    \includegraphics[width=0.65\linewidth]{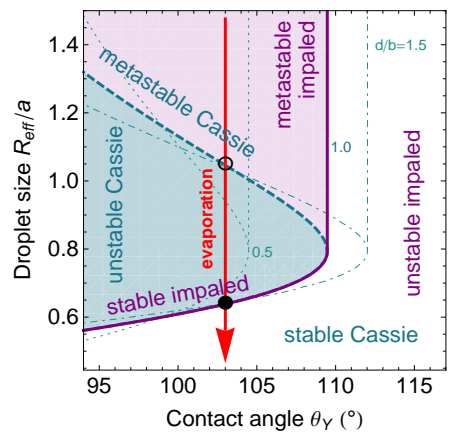}
    \caption{
    \textit{Phase diagram:} Regions of stability for the Cassie and the impaled state (intermediate minimum) as predicted by the analytical model. Focusing on the case of $d/b=1$ (thick lines), the impaled state is expected to exist in the complete shaded region, while the Cassie state is predicted to be \mbox{(meta-)stable} right to the dashed curve. An unstable impaled state (white region) automatically implies an absolutely stable Cassie state. Similarly, a unstable Cassie state goes along with an absolutely stable impaled state. In the lighter shaded area, a coexistence of metastable Cassie and impaled states is predicted. Thin dotted and dot-dashed curves give the theoretical predictions for other values of $d/b$. The arrow illustrates the path of a quasi-statically evaporating droplet as it enters the region of an unstable Cassie state ($\circ$), and finally goes over to a stable Cassie state again ($\bullet$).
    }
    \label{fig:phase-diagr-th}
\end{figure}

Figure~\ref{fig:phase-diagr-th} presents a morphological phase diagram displaying the theoretically expected regions of existence for the Cassie and the impaled state. The Cassie state is (meta-)stable for the set of points $(\theta_Y,R\eff)$ that fulfill $\upd f / \upd p|_{p=0}>0$, while the phase boundary for the impaled state is determined from $\upd f / \upd p=0$ combined with $\upd^2 f / \upd p^2>0$. Note that below a certain droplet size, the local minimum of the free energy shifts continuously to $p=0$ (Fig.~\ref{fig:model-predict}b), hence the impaled state now effectively appears as a Cassie state and the phase boundaries for the impaled and Cassie state are identical. In that case, the Cassie state becomes the only possible state (disregarding the Wenzel state).

Interestingly, the stability region of the Cassie state shows a characteristic shape, which is also largely independent of the surface geometry: There exists a certain droplet size where the Cassie state is unstable for a maximal range in contact angle, and both towards larger and smaller radii the stability region increases.

Noting that in the phase diagram a quasi-statically \footnote{Here, quasi-static refers to a sufficiently slow evaporation, such that the droplet assumes its optimum shape at any instant in time.} evaporating droplet would move on a vertical line from large towards small $R\eff$, we infer the existence of a \mbox{\textit{reentrant transition}}: A droplet, initially existing in a (meta-)stable Cassie state, can (following the arrow in Fig.~\ref{fig:phase-diagr-th}) become unstable upon a reduction of its size and thereupon adopt an impaled state ($\circ$). However, due to the fact that the position of the local free energy minimum shifts towards smaller $p$ with decreasing volume (Fig.~\ref{fig:model-predict}b), further evaporation will always result in the droplet to re-appear in a stable Cassie state ($\bullet$).

\section{Simulation results}

\begin{figure*}[t]
    (a) \includegraphics[width=0.26\linewidth]{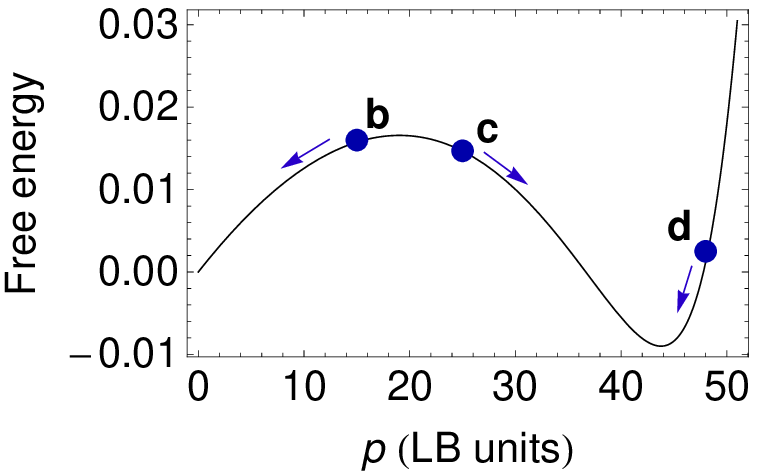} \
    (b) \includegraphics[width=0.18\linewidth]{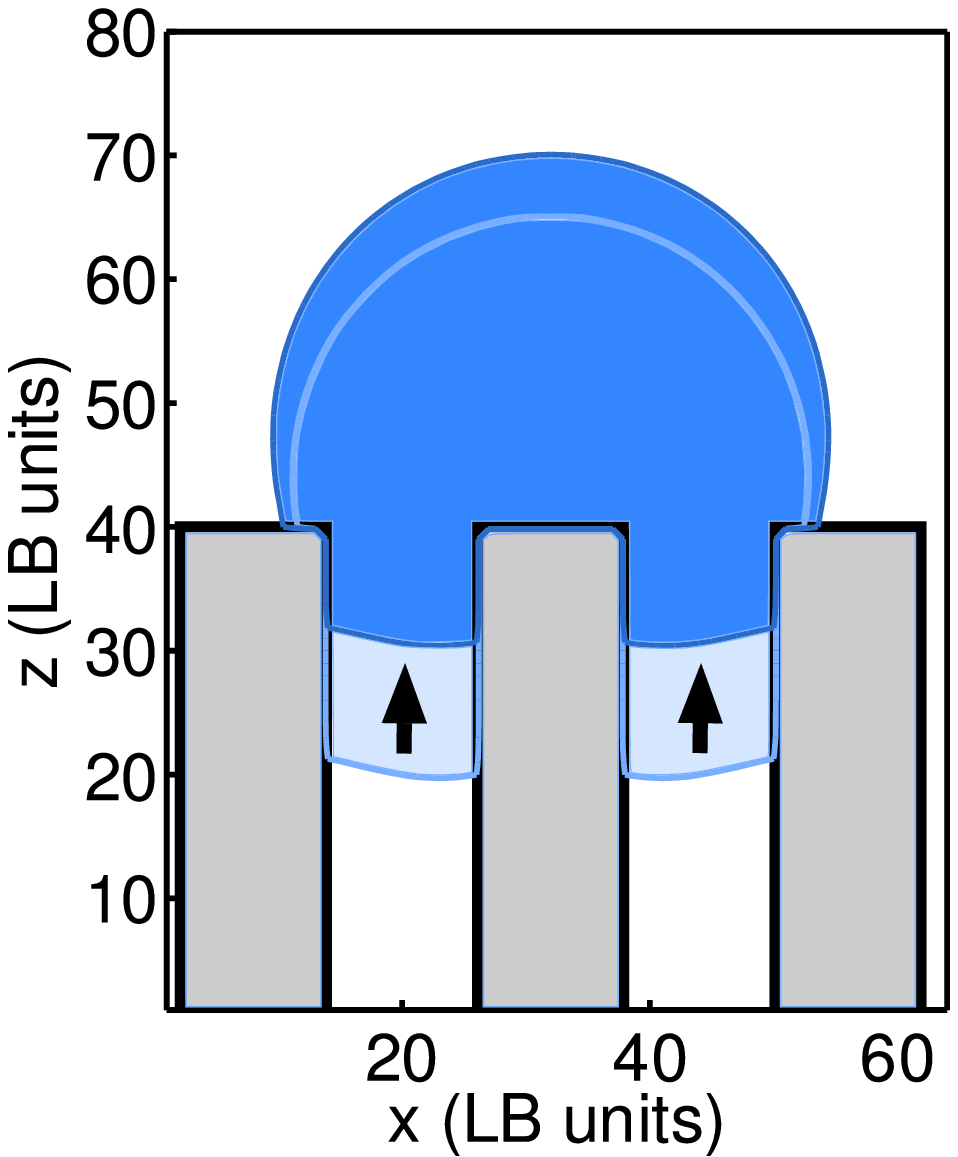} \
    (c) \includegraphics[width=0.18\linewidth]{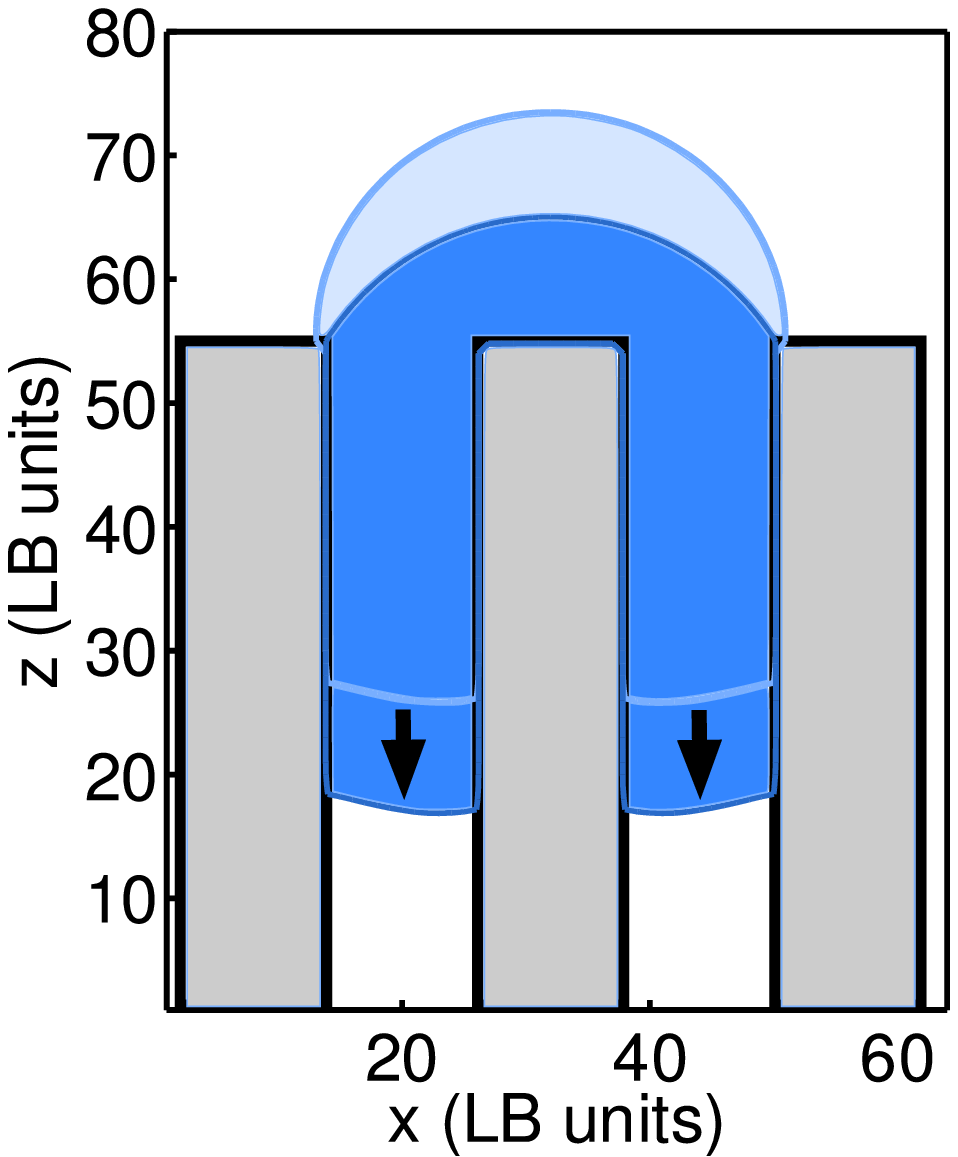} \  (d)\includegraphics[width=0.18\linewidth]{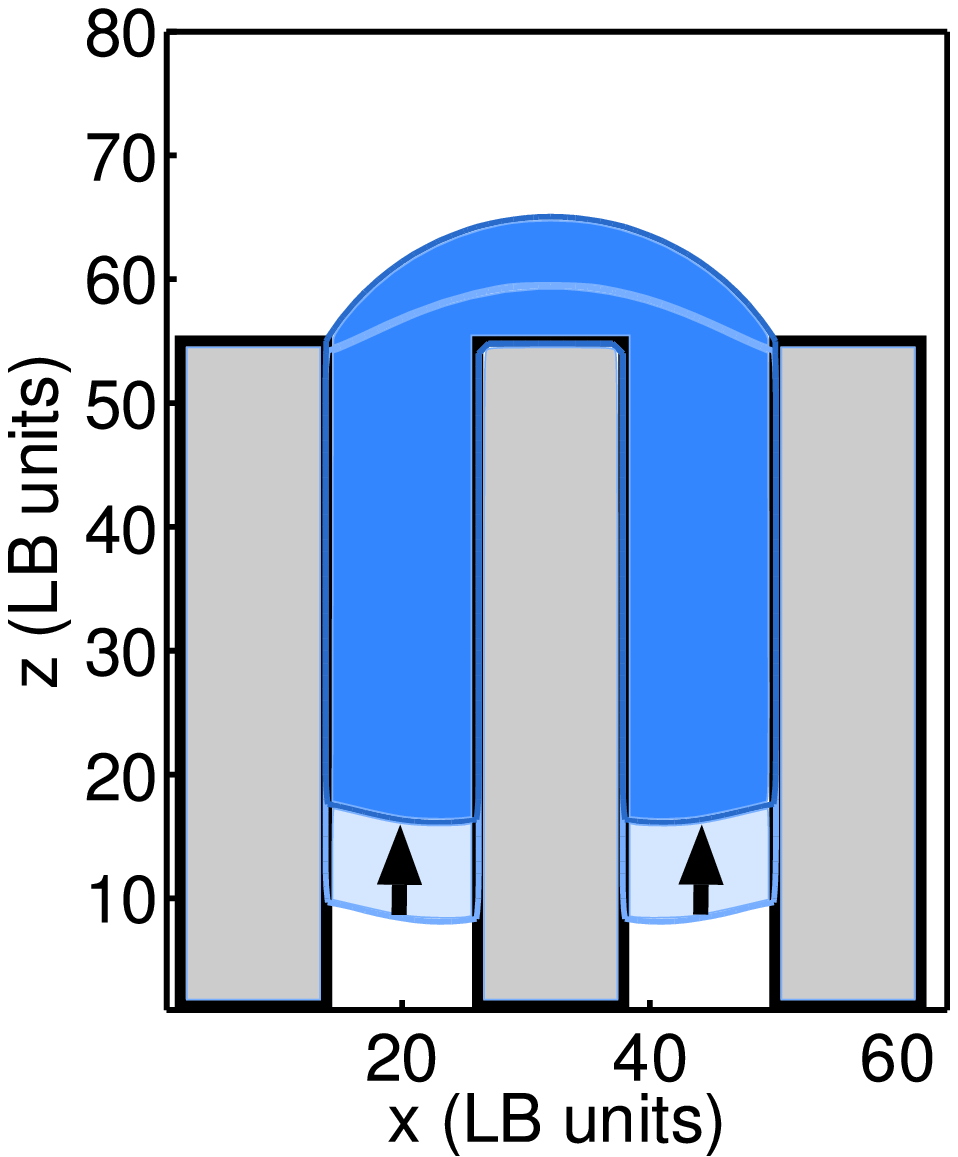}
    \caption{Droplet behaviour for different initial configurations. We take a fixed droplet size of $R\eff=1.2a$, a contact angle of $\theta_Y=103\deg$ and $b=d=12$\;\textnormal{LB units}. (a) shows the theoretical free energy $f/4\pi a^2 \sLV$ in dependence of the penetration depth $p$. In (b-d), the results of the LB simulations are displayed. The labeled dots in (a) mark the initial penetration depth of the droplet in the cases (b-d). The arrows indicate the predicted time evolution. The (partially impaled) Cassie state is observed in (b), while new impaled state is found in (c,d).}
    \label{fig:sim-meta}
\end{figure*}

We now compare the predictions of the analytical model to computer simulations. In contrast to standard single phase LB models \cite{succi_book, raabe_overview_2004, varnik_roughness_2007}, we employ here a free energy based two-phase (liquid-vapor) LB approach \cite{swift_lattice_1995, briant_lattice_2004}. Details of the algorithm can be found in \cite{dupuis_modelling_2005, varnik_wetting_2008}. The relaxation time is fixed to $\tau=0.8$. The temperature is set to $T=0.4$\;LB units, giving rise to stable coexisting liquid and vapour densities of $\rho_L\approx 4.1$ and $\rho_V\approx 2.9$, respectively. Note that this rather small density ratio (a well-known limitation of the present LB model) is not expected to adversely affect our results, since we are only concerned with the quasi-static (thermal equilibrium) behavior of a droplet.
For the interface parameter $\kappa$ we use a value of $0.002$, leading to an interface width of about 4 lattice nodes.
The simulation box mostly consists of $L_x\times L_y\times L_z=64\times 64\times 64$ lattice nodes, but is enlarged appropriately for the smaller droplets.
The substrate at the top ($z=L_z$) of the simulation box is flat. At the bottom ($z=0$), it is decorated with an array of equidistant cuboidal pillars. Typically, we will use $b=12$, $d=12$ and $h$ around 30 lattice units.
Periodic boundary conditions are applied along the $x$ and $y$ directions.

The relation between LB and physical units \cite{dupuis_modelling_2005} (assuming the simulated liquid is some viscous silicon oil) shows that the capillary time in our simulations is $t_c = 8\cdot 10^3\;\text{LB time steps} \approx 4\times 10^{-5}\;\text{s}$ and our droplets would be of micron scale. However, since it can be argued that the actual value of the physical viscosity is not important in the present case, it can be used to tune the unit of length, allowing the simulation results to be applied to a broad range of length scales, as long as thermal fluctuations and gravity can be disregarded.

We first of all demonstrate the metastability of the different wetting states, thereby establishing also the existence of the new impaled equilibrium state.
Indeed, as predicted by the corresponding theoretical free energy curve (Fig.~\ref{fig:sim-meta}a), the equilibrium position of a droplet depends on where it is placed at the beginning: A droplet deposited close to the top of the pillar array moves further to the top (Fig.~\ref{fig:sim-meta}b), while a droplet that initially penetrates deeper into the grooves becomes trapped in the intermediate minimum of the free energy (Figs.~\ref{fig:sim-meta}c,d) \footnote{Note that the free energy curves are just approximate descriptions of the real droplet behavior and, for example, do not predict the residual penetration depth of a droplet in the Cassie state.}.
This figure also nicely shows that the impaled state reported here (Figs.~\ref{fig:sim-meta}c,d) is indeed different from a partially impaled Cassie state (Fig.~\ref{fig:sim-meta}b).

\begin{figure*}[t]
    (a) \includegraphics[width=0.13\linewidth]{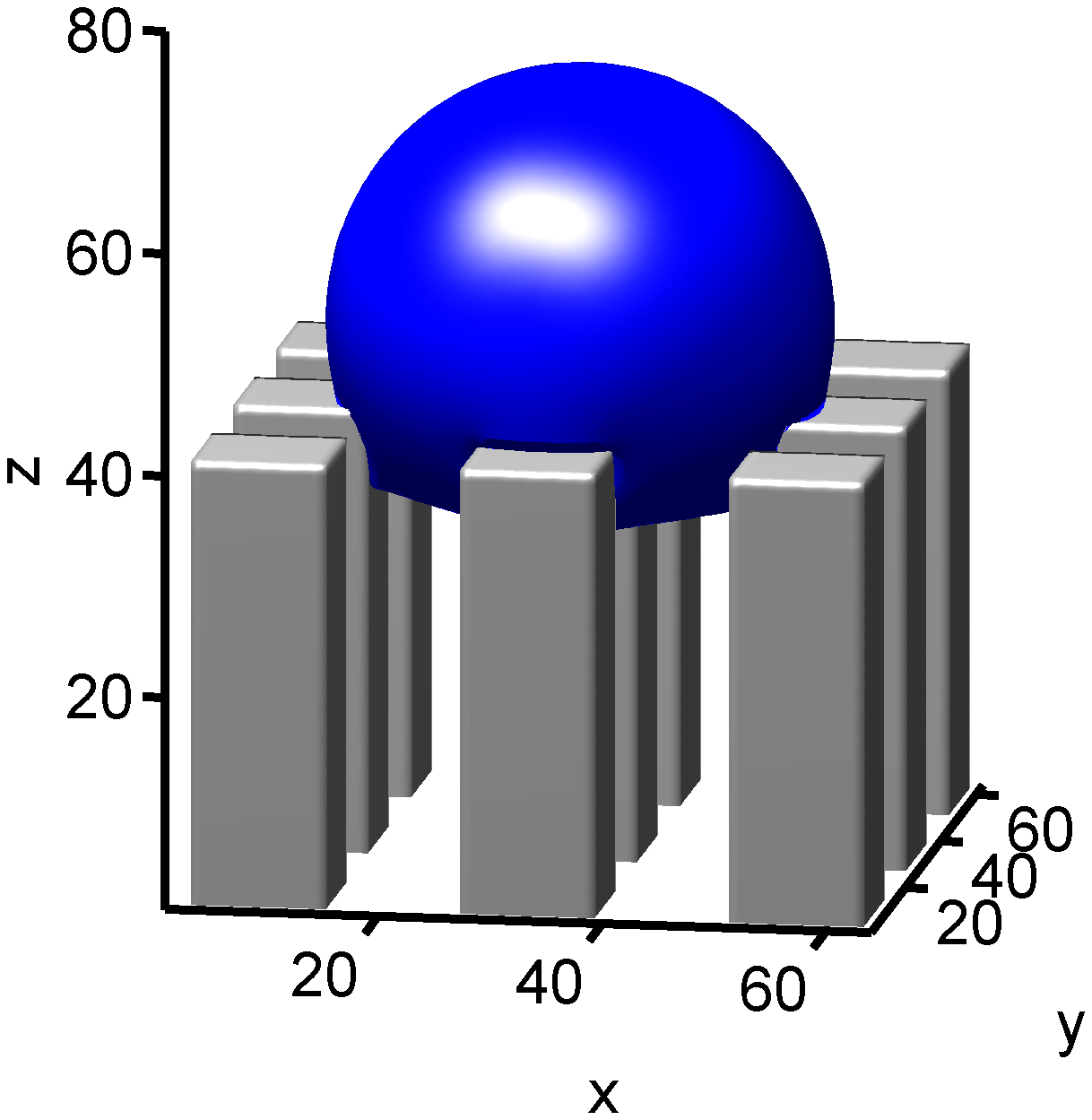} \;
    (b) \includegraphics[width=0.13\linewidth]{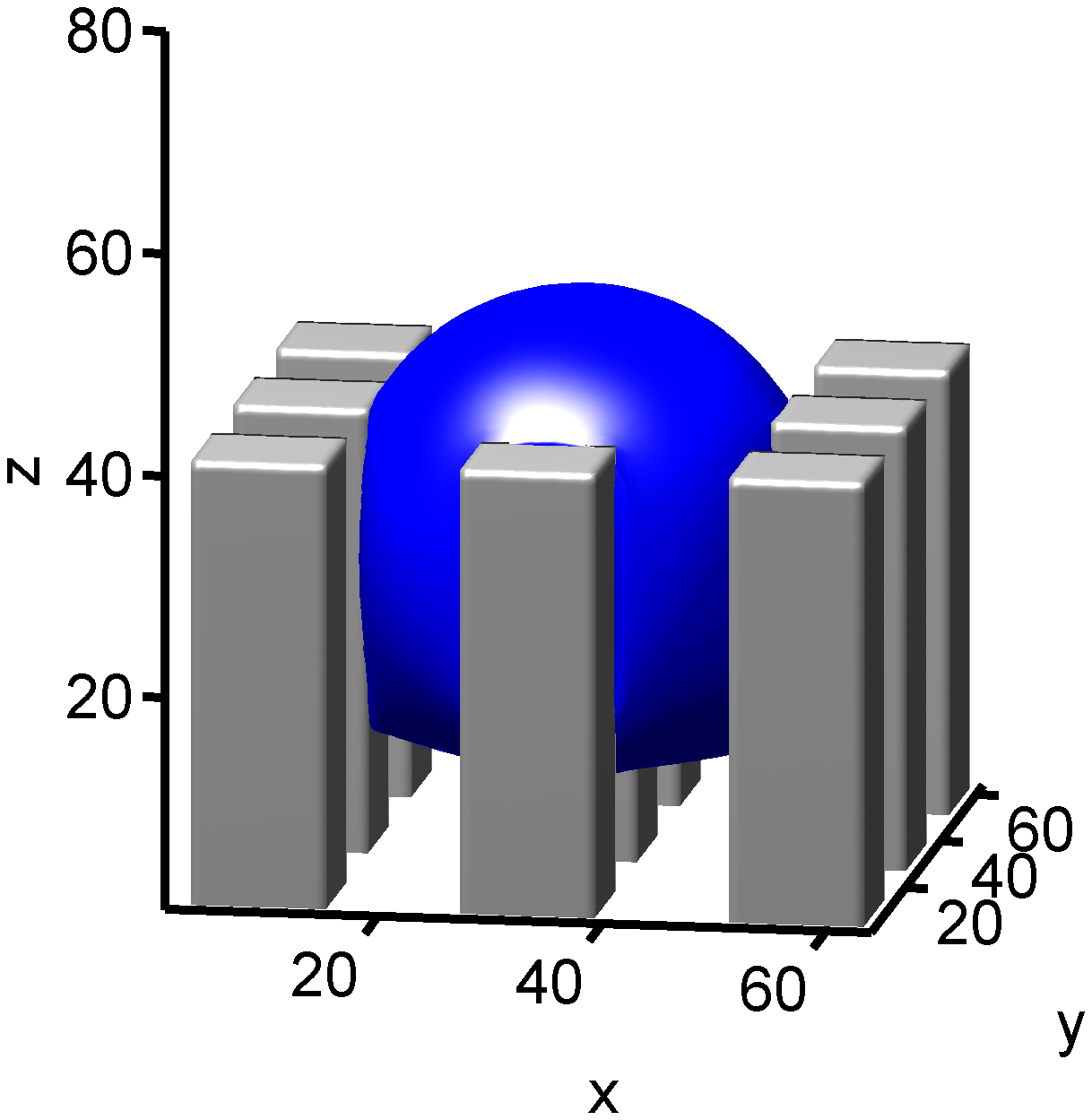} \;
    (c) \includegraphics[width=0.13\linewidth]{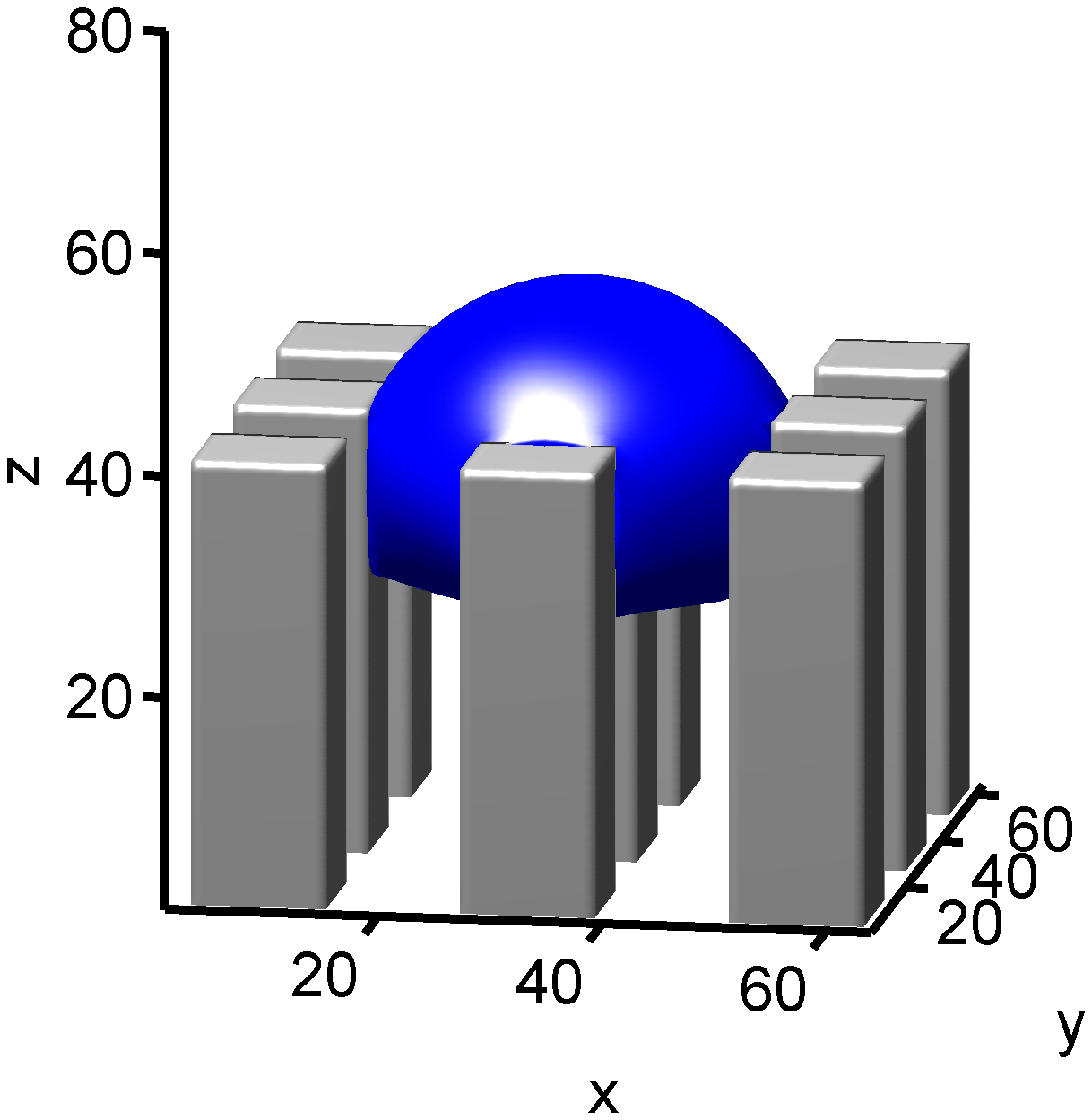} \;
    (d) \includegraphics[width=0.13\linewidth]{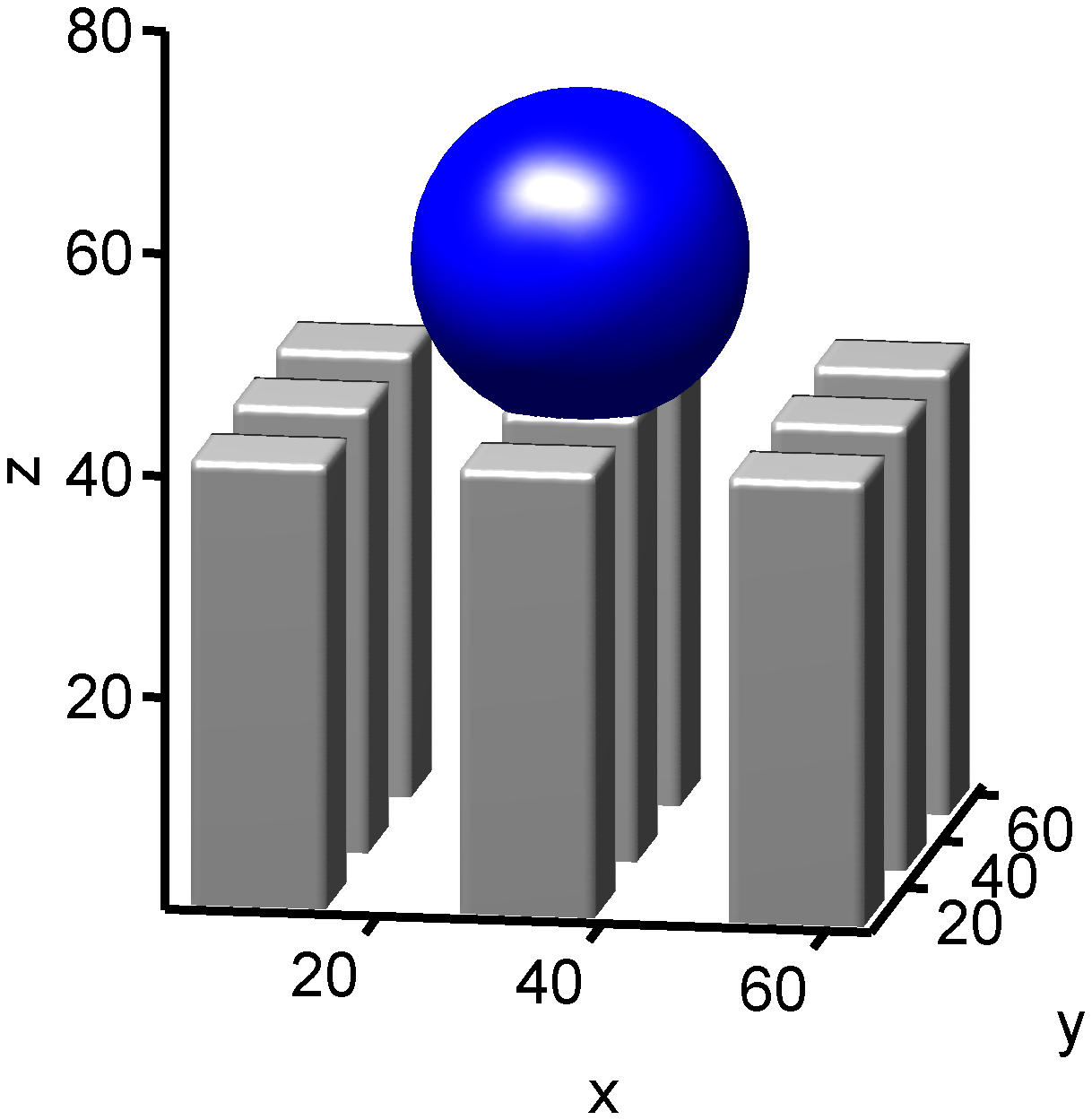}\;
    (e) \includegraphics[width=0.24\linewidth]{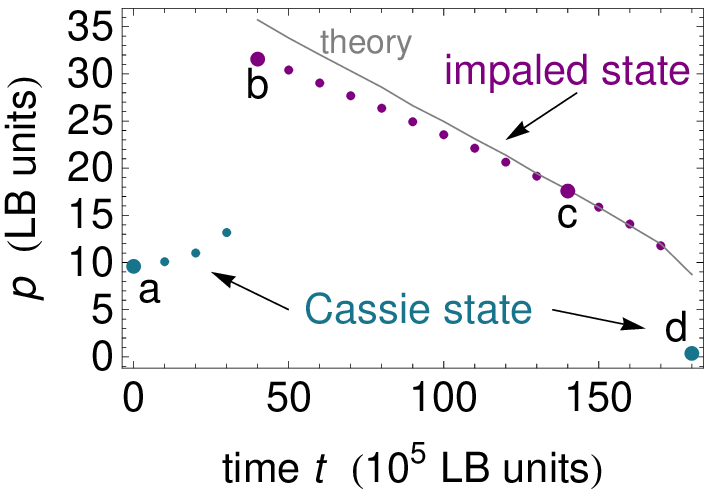}
    \caption{Reentrant transition through evaporation of a droplet. In the simulation, the droplet is initially (t=0) prepared in a (meta-)stable (partially impaled) Cassie state  (a). Evaporation is then switched on and proceeds by reducing, at a sufficiently low rate satisfying quasi-static equilibrium, the mass of the vapor phase across a $xy-$plane close to the top of the simulation box. In the course of the evaporation process, the Cassie state becomes unstable and adopts an impaled state (b), from where it gradually climbs up the pillars (c) until it reaches again a Cassie state (d). (e) shows the penetration depth of the lower droplet interface as it is observed in the simulation ($\bullet$) and predicted by the analytical model (---). Simulation parameters: contact angle $\theta_Y=103\deg$, initial droplet size $R\eff=1.2 a$, $b=d=12$\;\textnormal{LB units}, evaporation rate $5\times 10^{-7}$/\mbox{LB time steps}. All lengths are given in units of the LB grid spacing.}
    \label{fig:evap}
\end{figure*}

Figure~\ref{fig:evap} shows a simulation of the reentrant transition, achieved through a quasi-static evaporation process.
In the beginning (Fig.~\ref{fig:evap}a), the droplet resides in a (``partially impaled'') Cassie state and only slightly increases its penetration depth (thus confirming \cite{moulinet_life_2007, kusumaatmaja_collapse_2008}).  However, after reaching a critical size, the droplet suddenly penetrates into the pillar grooves and goes over to the intermediate minimum of the free energy (Fig.~\ref{fig:evap}b).
During the impaled phase (Figs.~\ref{fig:evap}b,c) the droplet gradually climbs up the pillars again, still residing in the local minimum. Note that its penetration depth is in nice agreement with the analytical predictions (Fig.~\ref{fig:evap}e). 
The final transition from the impaled to the Cassie state (Fig.~\ref{fig:evap}c,d) is, in contrast to the predictions of the analytical model (Fig.~\ref{fig:model-predict}), not continuous, but happens with the droplet depinning of from the outer pillars.

According to the common understanding of self-cleaning, impalement is considered unfavorable and the droplet cleans the surface by rolling over the top of the texture. Interestingly, the existence of a reentrant transition suggests the possibility of a qualitatively new self-cleaning mechanism, since the droplet not only touches the top of the substrate, but also its inner parts.

The existence of a reentrant transition can also explain some recent experimental observations, that found that small evaporating droplets indeed tend to remain close to the top of the substrate \cite{reyssat_impalement_2008, jung_wetting_2007} and not get trapped inside of the texture.


\begin{figure}[t]
    \centering
    \includegraphics[width=0.69\linewidth]{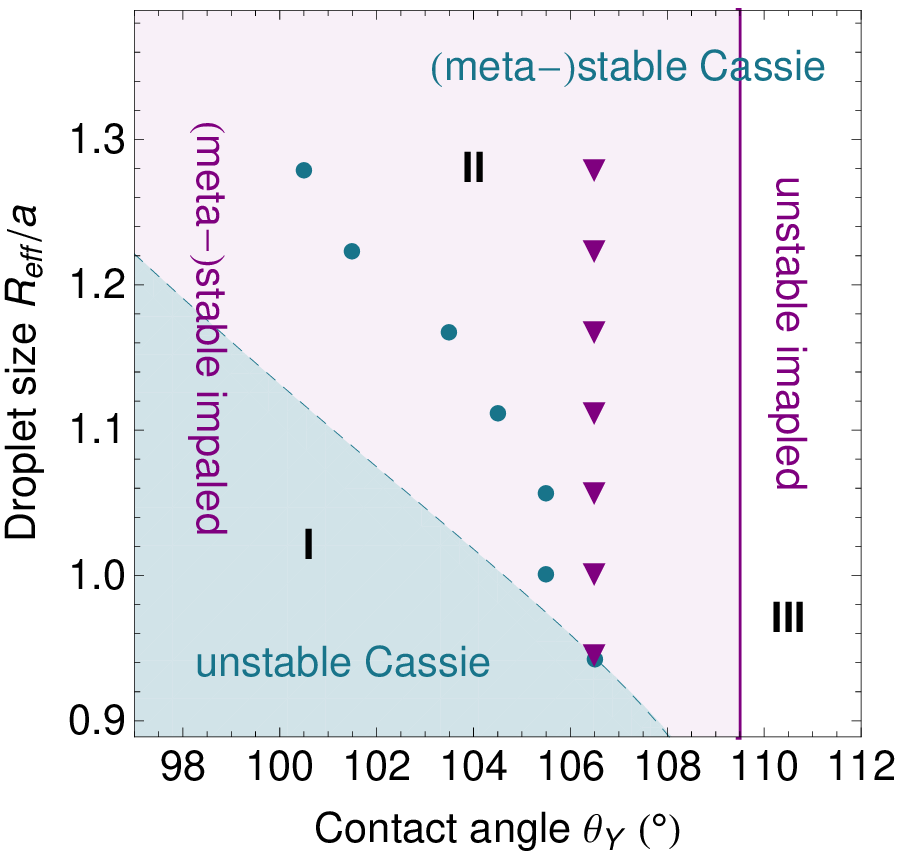}
    \caption{Stable droplet configurations in dependence of the droplet size and contact angle. In the analytical model, the Cassie state is expected to be unstable in region I (left to the dashed line), and \mbox{(meta-)stable} in regions II and III (right to the dashed line). The intermediate minimum of the free energy is predicted to exist in regions I and II (i.e.\ left to the solid line). Symbols ($\bullet,\blacktriangledown$) depict the phase boundaries as they result from our numerical simulations. There, the Cassie state is found to be unstable left to the $\bullet$. Similarly, the impaled state is found to exist left to the $\blacktriangledown$. Although deviating by a few degrees, both the numerical simulations and the approximate analytical model predict that the condition for the existence of the impaled state is independent of the droplet size.}
\label{fig:phase-diagr-sim}
\end{figure}
In Figure~\ref{fig:phase-diagr-sim}, the stability regions of the different droplet states predicted by the analytical model are investigated more closely. Due to limited computational power, only a part of the full phase diagram (Fig.~\ref{fig:phase-diagr-th}) is covered in the present case. The Wenzel state is also again not considered explicitly. Hence, for small contact angles and droplet sizes, the impaled state is expected to be the only stable state (region I). Conversely, for large contact angles, the Cassie state should be the only stable state (region III). Between these two extremes, the model predicts a region where \textit{both} a (meta-)stable Cassie and impaled state exist (region II).

As shown in Figure~\ref{fig:phase-diagr-sim}, despite its approximate nature, the overall droplet behaviour is described correctly by the analytical model and the three different regions are indeed recovered in the LB simulations.
The deviations between the simulated results and the theoretical phase diagram are not surprising, once the simplifications of the analytical model are taken into account. In particular the assumption of pinning at the inner pillar edges will be violated for larger droplets that are close to the top of the texture. They are observed to spread laterally into the grooves (increase their base radius) in order to fulfill the Young condition (cf. Figs.~\ref{fig:sim-meta}b and \ref{fig:evap}a).

The boundary for the existence of the \textit{intermediate minimum} of the free energy is found to be independent of the droplet size, as predicted by the analytical model. However, this line lies at $\theta_Y\approx 106.5\deg$, instead of the theoretically expected value of $\theta_Y\simeq 109.5\deg$. This deviation arises from the approximation of the impaled part of the droplet as a cylinder, which actually overestimates the contribution of this part to the total droplet volume. In the case of much larger droplets, contacting many pillars, the arguments given in our work suggest that an impaled state can always be realized provided that a contact between the droplet and the bottom substrate is inhibited.

Based on the Wenzel-Cassie model it can be shown \cite{bico_wetting_2002} that the free energy of a droplet is smaller in the Cassie state if 
$r > (\phi\cos\theta_Y-(1-\phi)) / \cos\theta_Y\,$. 
Here, $\phi$ is the pillar density (ratio of area covered by pillars and total projected area), $r$ is the surface roughness (ratio between entire surface area that can be wetted and its horizontal projection). For the present setup we have $r=3.5$ and $\phi=0.25$, leading to a value of $\theta_Y\simeq 103\deg$ below which the Cassie state would be unstable, regardless of the droplet size. As can be seem from Fig.~\ref{fig:phase-diagr-sim}, this estimate is clearly too crude for the small droplets considered here. Besides that, it also neglects possible free energy barriers between the Cassie and Wenzel state, which, as our work demonstrates, are crucial to the droplet phenomenology.

\section{Summary}

In conclusion, via analytical calculations and numerical simulations, we have uncovered the existence of an additional local equilibrium state, where the droplet partially wets the inside of the grooves, yet is not touching the base of the substrate. 
This finding qualitatively modifies the two-state paradigm of Wenzel and Cassie states in the case of droplets of comparable size to the surface roughness.
Interestingly, droplets in this new state appear to have Wenzel-like properties (e.g.\ small apparent contact angles)  -- but on the other hand possess the inherent capability to reenter the Cassie state again. We have demonstrated that our results are largely independent of the particular surface geometry and are expected to hold whenever a hydrophobic capillary is wetted by a small droplet. 

The insights presented here can serve as a valuable guidance for the fabrication of surfaces with specific wetting properties. Since we have explicitely shown that there is a maximal depth that a droplet can penetrate into the substrate, the optimal geometry of a surface can be easily assessed, depending whether complete (Wenzel) or incomplete wetting (Cassie) is desired.
Furthermore, our results are expected to contribute to a better understanding of how many surfaces occurring in nature can remain so perfectly clean and dry.

It is important to stress the scale invariance of our results (see Eq.~\eqref{eq:freeE}). 
This may, firstly, significantly widen the range of possible applications, and secondly, considerably simplify experimental verification of our predictions. One could e.g.\ study sub-millimetric drops on pillars of comparable size, thereby avoiding problems such as preparation and fast evaporation of micron-sized droplets.

\acknowledgments
We thank David Qu\'{e}r\'{e} who raised our interest in this topic and Alexandre Dupuis for providing us a version of his LB code. Financial supports by the Deutsche Forschungsgemeinschaft (DFG) under the grant numbers Va205/3-2 and Va205/3-3 (within the Priority Program Nano- \& Microfluidics SPP1164) as well as from the industrial sponsors of the ICAMS, the state of North Rhine-Westphalia and European Union are gratefully acknowledged.


\end{document}